\documentclass[conference]{IEEEtran}
\IEEEoverridecommandlockouts
\usepackage{cite}
\usepackage{amsmath,amssymb,amsfonts}
\makeatletter
\let\MYcaption\@makecaption
\makeatother

\usepackage[font=footnotesize]{subcaption}
\usepackage{tikz}

\usetikzlibrary{shapes,arrows,positioning,patterns}

\makeatletter
\let\@makecaption\MYcaption
\makeatother
\usepackage{algorithmic}
\usepackage{textcomp}
\def\BibTeX{{\rm B\kern-.05em{\sc i\kern-.025em b}\kern-.08em
    T\kern-.1667em\lower.7ex\hbox{E}\kern-.125emX}}

\usepackage{amsmath,graphicx,adjustbox}
\usepackage[utf8]{inputenc}
\usepackage[colorinlistoftodos]{todonotes}
\usepackage{siunitx}
\usepackage{placeins}
\usepackage{stfloats}
\usepackage{comment}

\usepackage{booktabs} 
\newcommand{\ra}[1]{\renewcommand{\arraystretch}{#1}}

\begin{document}
\title{Audio Impairment Recognition using a Correlation-Based Feature Representation\\  
\thanks{This publication has emanated from research conducted with the financial support of Science Foundation Ireland (SFI) under Grant Number 17/RC-PhD/3483 and 17/RC/2289\_P2 and was supported by The Alan Turing Institute under the EPSRC grant EP/N510129/1. EB is supported by RAEng Research Fellowship RF/128 and a Turing Fellowship.}
}

\author{\IEEEauthorblockN{Alessandro Ragano$^{1,2,3,4}$, Emmanouil Benetos$^{3,4}$, and Andrew Hines$^{1,2}$}
\IEEEauthorblockA{\textit{$^{1}$ School of Computer Science, University College Dublin, Ireland \ \ \ $^2$ Insight Centre for Data Analytics, Ireland}
\\ \textit{$^3$ School of EECS, Queen Mary University of London, UK \ \ \ $^4$ The Alan Turing Institute, UK}\\
alessandro.ragano@ucdconnect.ie, emmanouil.benetos@qmul.ac.uk, andrew.hines@ucd.ie
}
}

\maketitle

\begin{abstract}
Audio impairment recognition is based on finding noise in audio files and categorising the impairment type. Recently, significant performance improvement has been obtained thanks to the usage of advanced deep learning models. However, feature robustness is still an unresolved issue and it is one of the main reasons why we need powerful deep learning architectures. In the presence of a variety of musical styles, hand-crafted features are less efficient in capturing audio degradation characteristics and they are prone to failure when recognising audio impairments and could mistakenly learn musical concepts rather than impairment types. In this paper, we propose a new representation of hand-crafted features that is based on the correlation of feature pairs. We experimentally compare the proposed correlation-based feature representation with a typical raw feature representation used in machine learning and we show superior performance in terms of compact feature dimensionality and improved computational speed in the test stage whilst achieving comparable accuracy. 
\end{abstract}

\begin{tikzpicture}[overlay, remember picture]
\path (current page.north) node (anchor) {};
\end{tikzpicture}

\section{Introduction}
Audio classification tasks range from general classification of audio content e.g., music, speech or noise~\cite{dhanalakshmi2011classification}, to more specific classification of musical content e.g., music genre recognition or music annotation~\cite{Fu2011}. Identification and recognition of audio impairments is needed in several applications such as audio restoration of sound archives~\cite{Stallmann2016,Brandt2017}, voice activity detection~\cite{saeedi2015robust} and audio quality assessment~\cite{Reddy2019}.
Particularly, recognition of audio impairments can be useful in non-intrusive audio quality assessment i.e., when quality is predicted without the usage of the clean signal as opposed to intrusive quality metrics that make use of both clean and degraded signals. Non-intrusive quality assessment is more difficult to tackle and usually performs worse than the intrusive setting. However, non-intrusive metrics are necessary when assessing quality in real-time applications such as VoIP calls~\cite{hines2015measuring,hines2013monitoring} or when the clean signal cannot be available and sound is inherently noisy, for example with audio archives~\cite{Ragano2019}. Improving performance in the above-mentioned applications is critical for multimedia service providers in order to provide better quality of experience (QoE) for the end-users. 

In a typical machine learning scenario, researchers employ a two-stage approach: 1) extracting features, 2) designing the classifier. Feature extraction refers to the selection of salient and discriminatory measured values or the computation of secondary values from the measured values. Where the features are carefully selected or chosen for the task at hand they are often referred to as \emph{hand-crafted}. This selection process requires expertise or knowledge about the task and signal type under evaluation. 

For music classification, hand-crafted features include low-level features and high-level features. Low-level features such as spectral centroid, spectral bandwidth and zero crossing rate~\cite{Lerch2012} are typically extracted with a frame duration between 10 \si{ms} and 100 \si{ms}~\cite{Fu2011}. These features do not represent a human-level understanding of musical events, in contrast to high-level features such as pitch and beat which are closer to the perception of musical events. 

Hand-crafted features have been employed in audio classification tasks for noise identification~\cite{Brandt2017}, audio impairment recognition~\cite{Reddy2019,Brandt2017}, audio anomaly detection~\cite{Marchi2017} and general classification tasks in music informatics~\cite{Humphrey2013}. However, the problem of lack of robustness of those features is still an unresolved issue. Improved performance in music informatics tasks is mainly due to advanced deep learning models which partly compensate the lack of robustness in hand-crafted features as explained by Humphrey et al.~\cite{Humphrey2013}. This problem is emphasized in audio impairment recognition and noise identification scenarios where hand-crafted features could fail in recognising impairment types given the presence of different styles of musical content. 

In this paper, we propose a new representation of hand-crafted features based on the correlation between pairs of features for the tasks of audio impairment recognition and noise identification. Rather than feeding the classifier with raw feature values we use a correlation-based feature representation as input of the classifier. We carry out experiments on noise identification and audio impairment recognition. To demonstrate the wide applicability of the proposed method we also investigate the music genre recognition task which is widely explored in music informatics. Insights from our study show that the correlation-based feature representation achieves the same accuracy of hand-crafted raw features and results in reduced feature dimensionality and higher computational speed when testing the model. 

\section{Related Work and Motivations}
The motivation to explore the effect of using the correlation between features as input of the classifier originated while exploring the problem of feature robustness in the presence of noise for audio quality assessment applications~\cite{Akhtar2017,Ragano2019}. Classifying the impairment type can be useful when predicting audio quality in non-intrusive scenarios as it can give information on the expected perceived audio distortion \cite{Akhtar2017}. 
\begin{figure*}[t]
\centering
  \begin{subfigure}[c]{0.44\linewidth}
  \centering
    \includegraphics[width=\textwidth,keepaspectratio=True]{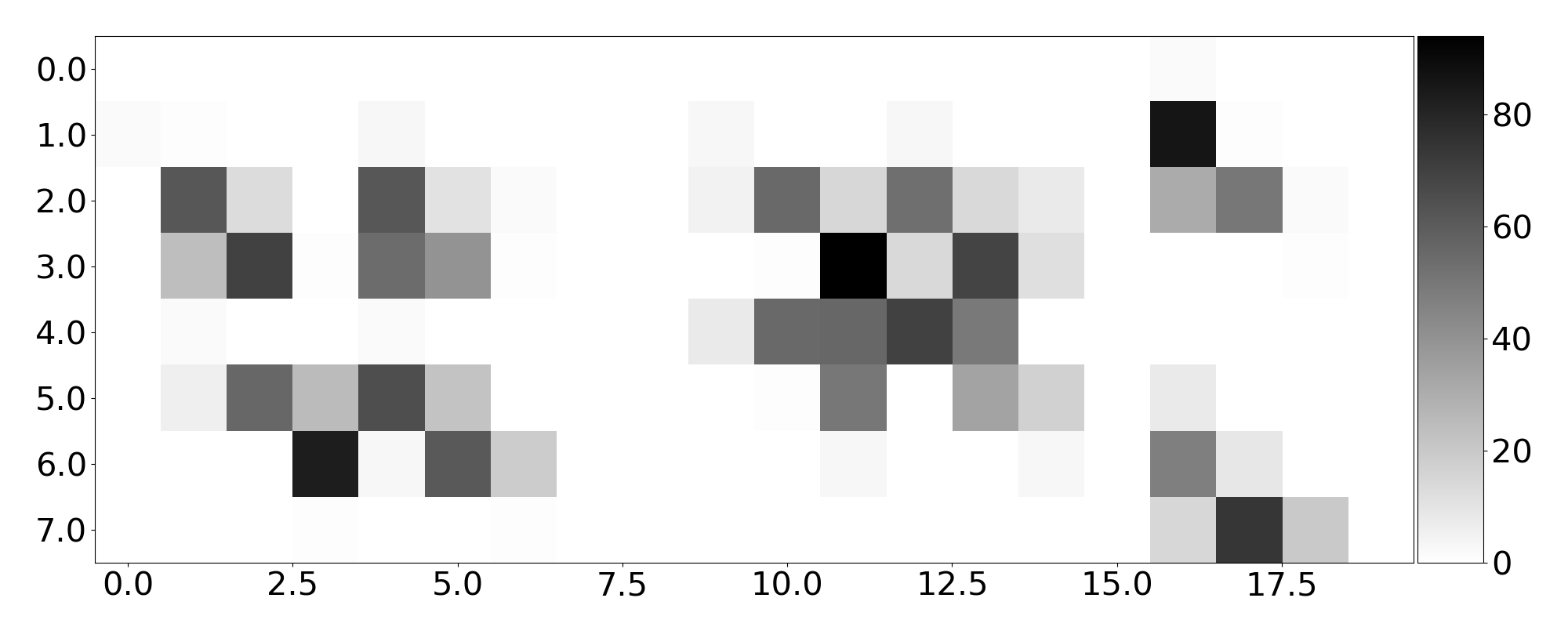}
     \caption{Combination of 10 two-dimensional histograms using 4 bins.}
  \end{subfigure}
  \quad
  \begin{subfigure}[c]{0.44\linewidth}
  \centering
    \includegraphics[width=\textwidth,keepaspectratio=True]{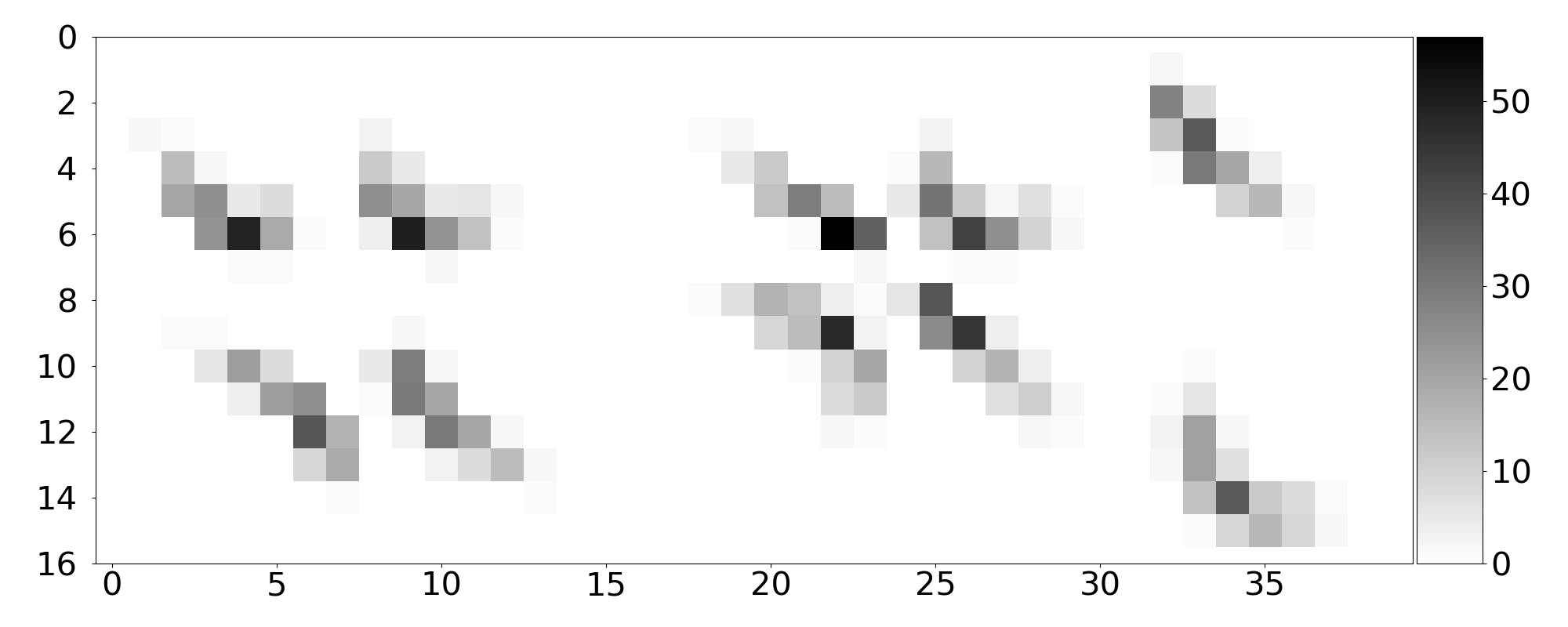}
    \caption{Combination of 10 two-dimensional histograms using 8 bins.}
  \end{subfigure}
  \\
  \begin{subfigure}[c]{0.44\linewidth}
  \centering
    \includegraphics[width=\textwidth,keepaspectratio=True]{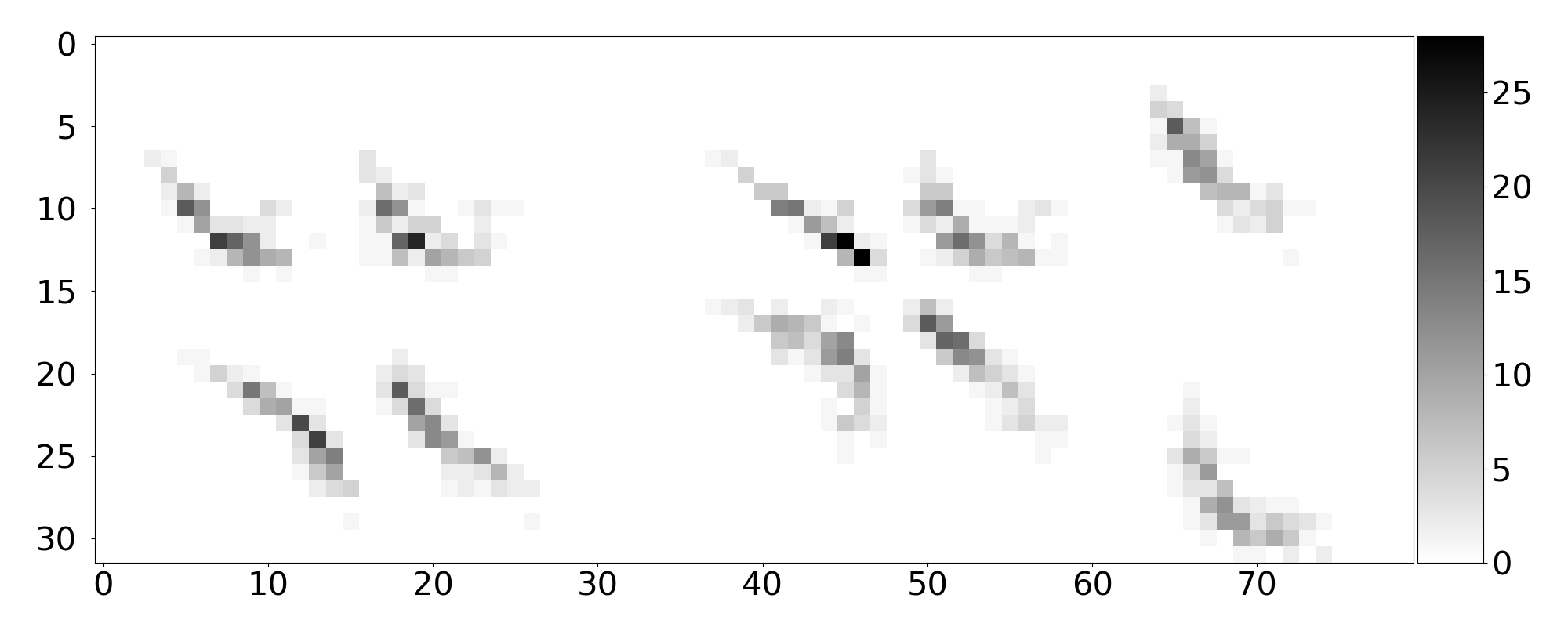}
     \caption{Combination of 10 two-dimensional histograms using 16 bins.}
  \end{subfigure}
  \quad
  \begin{subfigure}[c]{0.44\linewidth}
  \centering
    \includegraphics[width=0.90\textwidth,keepaspectratio=True]{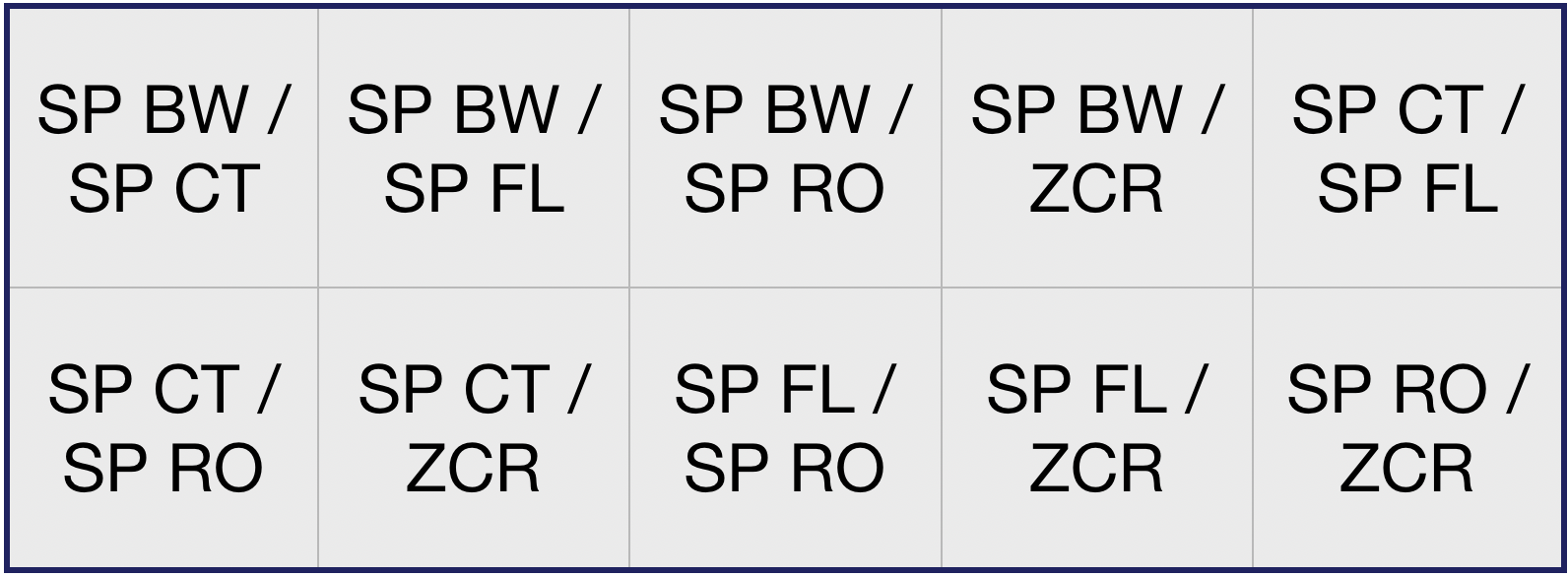}
    \vspace{3mm}
    \caption{Respective names of each feature pair.}
    \vspace{-2mm}

  \end{subfigure}
  \caption{An example of the correlation-based representation of one training sample belonging to the first 4 seconds of "The Slump" by Tony Williams mixed with a vinyl hiss at 30 dB. 4 bins representation (a), 8 bins representation (b), and 16 bins representation (c). The corresponding positions of the feature pairs are shown in (d).}  \label{fig:2D_histogram}
\end{figure*}

We assume that a more compact representation of features could be achieved by representing correlation between each pair of features rather than using raw feature values. In this way, our expectation is that we provide more informative features to the classifier by attenuating redundant information in hand-crafted features. New directions in music informatics propose the use of feature learning instead of using hand-crafted features \cite{Humphrey2013}. However, the same does not apply for audio impairment recognition where hand-crafted features and spectrograms are still employed \cite{Reddy2019}. For this reason, we decided to use hand-crafted features for our task.
Designing robust features for general music informatics tasks has been analysed by Humphrey et al.~\cite{Humphrey2013}. The authors discover that insufficient non-linearity, poorly tuned parameters, and the inherent problems with short-time analysis are the main problems of hand-crafted features. One of the main conclusions from their analysis is that the lack of robustness in features can be partly compensated by more powerful classifiers which are based on deep learning. The same problem is found in audio impairment recognition~\cite{Reddy2019, Brandt2017, Stallmann2016, saeedi2015robust}. The authors have shown that hand-crafted features are not robust in various contexts. They conclude that more powerful classifiers such as deep convolutional neural networks (CNNs) or features based on psychoacoustic models are needed to reach high performance. A similar scenario is also found in audio anomaly detection where detecting noise in the audio data requires advanced deep learning models to compensate for the lack of feature robustness~\cite{Rushe2019,Marchi2017,li2018anomalous}. Modifying feature representation has been successful when detecting non-speech audio events where methods such as principal component analysis (PCA) have been used for reducing dimensionality and improve feature robustness at the same time \cite{portelo2009non}. 
Motivated by the above-mentioned issues, in this paper we create a correlation-based representation to tackle the above-mentioned problem of feature robustness. Our assumption is that the correlation as a two-dimensional representation with spatial relationships could be more robust for detecting noise hidden in a dataset characterised by many different musical styles given that it could eliminate redundant information that belongs to the music content rather than the impairment type.
The goal of the paper is to contrast two feature representations in the presence of noise. 
We carry out 3 experiments: noise identification, audio impairment recognition, and music genre recognition. We assess the differences in accuracy, feature dimensionality, and running time between the correlation-based feature representation and the typical usage of raw feature values. Contrary with our assumptions, we show that there is no improvement in terms of feature robustness where both feature representations perform similarly. However, results show improvement in terms of feature dimensionality reduction and running time which support the usage of the proposed method. 

\section{Representation of Feature Correlation using Two-Dimensional Histograms}
The correlation-based feature representation can be obtained by generating the equivalent of the individual scatter plot used for analysing the correlation between every pair of features. 
To represent the correlation we generate an individual scatter plot by computing the two-dimensional histograms of each pair of features employed in the experiment.

The correlation-based representation is generated as follows:
\begin{enumerate}
 \item We split the whole audio track into frames of 4 seconds with 50\% of overlap as done in ~\cite{Senac2017, Zhang2016}. 
 
 \item Given an audio frame we compute hand-crafted features. We compute two groups of features $F$ as defined in~\cite{Lerch2012}. The first group, the short-time (ST) features, includes spectral roll-off (SP~RO), spectral centroid (SP~CT), zero crossing rate (ZCR), spectral bandwidth (SP~BW), and spectral flatness (SP FL). The second group is characterised by the first 5 mel-frequency cepstral coefficients (MFCCs).
 
 \item For each pair of features we compute a 2D histogram specifying the number of histogram bins  $B$ and the histogram range. The number of histogram bins is a crucial parameter as we will show in the results section.
 
 \item The histogram range is given by the maximum/minimum absolute value of each feature with respect to the whole dataset. In this way we avoid eventual similarity between different pairs of features due to the automatic scaling of each feature when combining different individual 2D histograms.
 
 \item We combine each individual 2D histogram of each pair of features in a bigger matrix which represents a training sample. We concatenate along the rows or the columns depending on the number of features employed.
 \end{enumerate}
Given that we use $F=5$ features we need $\binom{F}{2}=10$ individual two-dimensional histograms to be concatenated with each other. Each individual histogram has a size equal to $B \times B$. Therefore the size of one training sample is given by:
$$(2 \times B) \times (5 \times B)$$
where $(2 \times B)$ is the number of rows and $(5 \times B)$ is the number of columns. 
\begin{table*}\centering
\caption{CNN architecture used in Audio Impairment Recognition (AIR) and Noise Identification (NI) by using the Correlation and the Raw representation. We show the number of channels ($c$), the kernel size ($k$), and the padding factor $(p)$ for convolutional layers, the pooling size ($ps$) for the max pooling layers and the number of nodes ($n$) for the fully connected layers. $F=5$ is the number of employed features.}
\ra{1.3}
\begin{Huge}{}
\begin{adjustbox}{max width=0.90\textwidth}
\begin{tabular}{@{}llllllllllll@{}}\toprule

& \multicolumn{3}{c} {Audio Impairment Recognition}& \phantom{abc} & \multicolumn{4}{c}{Noise Identification} & \multicolumn{1}{c}{} & \phantom{abc} & \multicolumn{1}{c}{} \\
\cmidrule{2-4}  \cmidrule{6-12}

\textbf{Operation} &  AIR-Correlation && AIR-Raw && NI-Correlation &&  NI-Raw &  \\ \midrule
Conv & $c=16$, $k=(7,7)$, $p=same$  &&  $c=16$, $k=(4,F)$ && $c=8$, $k=(3,3)$, $p=same$ && $c=8$, $k=(4,F)$ \\
Conv & $c=32$, $k=(5,5)$ && $c=32$, $k=(4,1)$ && $c=16$, $k=(3,3)$ && $c=16$, $k=(4,1)$ \\  MaxPool[Dropout(50)] & $ps=(2,2)$ && $ps=(Prev Output,1)$ && $ps=(2,2)$ && $ps=(Prev Output,1)$ \\
Dense[Dropout(50] & $n=200$ && $n=200$ &&  $n=128$ &&  $n=128$\\
Dense & $n=4$ && $n=4$ && $n=2$ && $n=2$ \\ \bottomrule
\end{tabular}
\end{adjustbox}
\end{Huge}
\label{table:archi}
\end{table*}

In Figure~\ref{fig:2D_histogram} we visualise one training sample from our dataset which
has been created by mixing tracks from the GTZAN dataset ~\cite{Tzanetakis2002} with impairments taken from the Freesound database~\cite{Font2013}. The track is ``The Slump" by Tony Williams mixed with a vinyl hiss at 30 dB. We show the correlation-based feature representation using 4 bins, 8 bins, and 16 bins indicating the respective feature pairs. The number of histogram bins $B$ affects both resolution and dynamic range of the proposed feature representation. A small value of $B$ results in higher dynamic range but lower resolution. This is due to the fact that with using fewer histogram bins we associate more different points to the same square as shown in Figure \ref{fig:2D_histogram}a that consequently increases the magnitude at the expense of lower resolution. 

It should be noted that we experimented with using images of the scatter plot matrix generated from data analysis libraries such as Pandas as alternative to the proposed histogram-based method but informal tests yielded inferior results compared to the histogram method.  

\section{Experimental Setup}
\label{sec:majhead}
In this section we describe the experimental setup which includes a description of the dataset and model architectures.

\subsection{Dataset}
We employ the GTZAN dataset~\cite{Tzanetakis2002}, widely used for music genre recognition~\cite{Sturm2014}, because it allows us to use a good variety of musical styles avoiding any potential genre-specific bias. The dataset includes 1000 tracks equally divided in 10 different genres, each with 30 seconds duration. For audio impairment recognition and noise identification we extract 120 tracks from the dataset by excluding those containing repetitions, mislabelings and distortions as discussed by Sturm ~\cite{Sturm2013}. We mix the 120 tracks with different audio impairments as discussed below.

\subsection{Model Architectures}
In each experiment we use a CNN as classifier. The motivation to use CNNs is due to their success in computer vision applications where the input of the architecture is a two dimensional matrix with spatial relationships as in the case of the proposed method. In Table~\ref{table:archi} we show the architectures used in the two tasks for each group of features. In both tasks we try to keep a very simple architecture to minimise the capacity of the classifier. Regarding raw features we use a domain-knowledge approach ~\cite{Pons2017} to design the kernels. By using 5 columns in the kernel of the first convolutional layer, the model learns the combination of features. Then we design the second convolutional layer to model the temporal dependencies as adopted similarly in ~\cite{Senac2017}. We have noticed a more robust learning when using this approach. 

\section{Results}
The goal of these experiments is to compare the proposed feature representation with a typical feature representation that uses raw values. We evaluate performance by varying the histogram bins $B$ in the proposed method to assess the accuracy, the size of the training dataset, and running time for the test stage. The size is expressed in terms of feature dimensionality and represents the number of matrix entries in the training dataset. We expect that the running time is shorter when using a more compact representation. It should be noted that the dataset is created in both methods with the same amount of audio data and the size difference in the training data depends on the employed representation. The running time is the amount of time that the model takes to be tested and it is computed on a MacBook Pro with 6-Core Intel i9 2.9GHz. One training/test sample captures 4 seconds of audio with 2 seconds of overlap in both raw and correlation representations. This aggregation of features over time has been used to improve the classification accuracy as discussed in~\cite{Zhang2016}. We split the dataset into training, validation, and test subsets. The training and test subsets are partitioned using cross-validation while the validation set represents 15\% of the training dataset obtained after cross-validation. We use the validation set to find the best model i.e., the one with the lowest categorical cross-entropy loss after 800 epochs. The best model is selected for evaluation with the test set. It should be noted that the split is made on a track-level instead of splitting on a frame-level to avoid any possible repetitive use of the data between the training set, the test set, and the validation set.

\begin{table*}
\caption{Performance evaluation of Audio Impairment Recognition and Noise Identification with MFFCs and ST features. We compare the accuracy, the size, and running time at 3 different histogram bins $B$ between the proposed method Corr(B=4,B=8,B=16) and the raw feature representation. Bold text indicates the cases where the proposed method shows superior performance.}
\centering
\Large
\ra{1.3}
\begin{adjustbox}{max width=0.95\textwidth}
\begin{tabular}{@{}lclccclclclccccccll@{}}\toprule

& \multicolumn{7}{c} {Audio Impairment Recognition - MFCCs}& \phantom{abcabc} & \multicolumn{7}{c}{Audio Impairment Recognition - ST} & \multicolumn{1}{c}{} & \phantom{abc} & \multicolumn{1}{c}{} \\
\cmidrule{2-8}  \cmidrule{10-16}

&  Corr(B=4) && Corr(B=8) && Corr(B=16) && Raw && Corr(B=4) && Corr(B=8) && Corr(B=16) && Raw &  \\ \midrule
Accuracy & 57.49 &&  \textbf{63.57} && \textbf{63.45} && 63.05 && 54.64 && \textbf{59.35} && \textbf{61.78} && 59.16 \\
Size & \textbf{160} && \textbf{640} && 2560 && 865 && \textbf{160} && \textbf{640} && 2560 && 865  \\
Running Time (\textit{secs}) & \textbf{2.21} && \textbf{2.71} && 6.35 && 4.71 && \textbf{2.05} && \textbf{3.44} && 6.69 && 5.22\\
\bottomrule
\vspace{0.4cm}
\end{tabular}
\end{adjustbox}
\begin{adjustbox}{max width=0.95\textwidth}
\begin{tabular}{@{}lclccclclclccccccll@{}}\toprule

& \multicolumn{7}{c} {Noise Identification - MFCCs}& \phantom{abcabc} & \multicolumn{7}{c}{Noise Identification - ST} & \multicolumn{1}{c}{} & \phantom{abc} & \multicolumn{1}{c}{} \\
\cmidrule{2-8}  \cmidrule{10-16}

&  Corr(B=4) && Corr(B=8) && Corr(B=16) && Raw && Corr(B=4) && Corr(B=8) && Corr(B=16) && Raw &  \\ \midrule
Accuracy & 73.15 && 76.15 && 77.08 && 77.91 && 73.57 && 73.79 && 76.54 && 77.38 \\
Size & \textbf{160} && \textbf{640} && 2560 && 865 && \textbf{160} && \textbf{640} && 2560 && 865  \\
Running Time (\textit{secs}) & \textbf{2.40} && \textbf{3.09} && 6.35 && 4.90 && \textbf{2.30} && \textbf{3.36} && 7.09 && 4.70 \\
\bottomrule

\end{tabular}
\label{table:results}
\end{adjustbox}
\end{table*}

\subsection{Audio Impairment Recognition}
We perform a stratified 6-fold cross validation to test our model which guarantees a balanced trade-off between bias and variance and equal partition of the data with respect to the ground truth. We classify 4 types of noise that are typically found in archive recordings: vinyl hiss, tape noise, gramophone noise, and white noise. The first 3 real-world noises have been obtained from the Freesound database~\cite{Font2013}. We create mixtures at different SNR levels from 0 dB to 30 dB with 5 dB increments as done by Reddy et al.~\cite{Reddy2019}. The goal is to classify the 4 different impairments. The accuracy and runtime are averaged over the different test partitions while the size is not averaged since it is the same in every experiment. The results of the accuracy, the size, and running time are shown in Table~\ref{table:results}. The correlation-based feature representation with 8 bins allows to use a dataset $\approx30\%$ smaller than the one obtained from the raw representation. Regarding running time, the correlation representation allows to have a model which is $\approx53\%$ faster when using MFCCs and $\approx41\%$ faster with ST features. Shorter running time is due to the compact feature representation as expected.   
The accuracy of the classifier has no significant difference between the correlation representation and the raw values. The usage of other bin values is discouraged due to lower accuracy ($B=4$), larger dataset ($B=16$) and longer running time ($B=16$). The difference in the results between MFCCs and ST features is negligible when comparing the correlation representations with the raw values. However, it is significant in terms of absolute performance where MFCCs outperform ST features. The obtained accuracy scores show that the proposed method does not address the hypothesis of feature robustness that we discussed above. However, reduction in both size and running time support the usage of the proposed method for reducing feature dimensionality. 

\subsection{Noise Identification}
As with the previous task, we perform a stratified 6-fold cross-validation to test the proposed model in several equally split partitions. Here we have a binary classification problem. Half the dataset contains clean recordings while the other half contains mixtures at 30, 15, and 0 dB SNR levels. The types of noise employed are the same as above. The mixtures with SNR levels lower than 30 dB are considered as noisy recordings. Using different SNR levels guarantees more robustness during training and ensures that the classifier learns from the noise instead of the different mixture levels. The goal is to distinguish noisy recordings from the clean ones regardless of the noise type. We decided to include different types of noise to explore a harder task compared to the case with only one class of noise. The results are shown in Table~\ref{table:results}. We obtain results similar to the audio impairment recognition task. The correlation-based feature representation results in a dataset $\approx30\%$ smaller and in a model runtime $\approx43\%$ faster with MFCCs and $\approx33\%$ faster with ST features despite a comparable accuracy. Again, results support the usage of the proposed method for reducing feature dimensionality instead of improving feature robustness as we assumed in the motivations section above.

\subsection{Music Genre Recognition}
\label{sec:mgr}
In these experiments we apply a 10-fold cross validation on the whole GTZAN dataset as used in~\cite{Tzanetakis2002, Senac2017, Fu2011}. Unlike the previous tasks, we evaluate only the accuracy by setting the histogram bins equal to 8 as we are interested in understanding the capability of the method in a different task and using a much larger dataset. Therefore, we have a dataset that $\approx30\%$ smaller than the raw feature representation.

\begin{table}[t]
\ra{0.9}
\centering
\caption{Music genre recognition accuracy.}
\begin{Huge}{}
\begin{adjustbox}{max width=0.48\textwidth,center}
\begin{tabular}{>{\LARGE}l>{\LARGE}c>{\LARGE}c>{\LARGE}c} 
\toprule
 \textbf{Method} & \textbf{Features} & \textbf{Classifier} & \textbf{Accuracy} \\ [0.5ex] 
 \midrule
 ~\cite{Tzanetakis2002} & \{8 ST+12 MFCCs\}$\times$MuVar+beat+pitch & GMM & 61\% $\pm$ 4\% \\  
 ~\cite{Senac2017} & 5 ST+2 Tonality+ ST Energy & CNN+RB & 89.6\% $\pm$2.4\% \\
 Raw Features & 5 ST & CNN & 72.20\% $\pm$ 4.26\% \\
 Correlation Features & 5 ST & CNN & 68.90\% $\pm$ 4.54\% \\
 Late Fusion & 5 ST {Raw and Corr.} & CNN & 74.70\% $\pm$ 5.24\% \\ [1ex] 
\bottomrule
\end{tabular}
\end{adjustbox}
\end{Huge}
\label{table:mgr}
\end{table}
The goal of this task is to classify 10 genres as labelled in the GTZAN dataset. Results are shown in Table \ref{table:mgr}. The proposed method shows a slightly lower accuracy to the one using the raw feature representation with the advantage of $\approx30\%$ reduction in feature dimensionality. It should be noted that this method does not outperform the best model ~\cite{Senac2017} which uses CNNs with residual blocks (RB) and a combination between ST features and time-frequency energy. This result is aligned with the previous tasks where we show that the advantages of the proposed method are due to feature dimensionality and running time instead of higher accuracy. 
We also explore the combination of the raw feature values and the correlation after training the two models separately. The accuracy of 74.70\% suggests that the two methods do not learn exactly the same concepts.

\section{Discussion}
In this paper we have shown that using the correlation between features using two dimensional histograms allows us to use a reduced feature dimensionality and a shorter running time whilst achieving comparable accuracy. The performance of the proposed method depends on the choice of the histogram bins $B$ up to a limit of $16$ bins (Figure \ref{fig:2D_histogram}c). Testing with higher values for $B$ had no positive effect on the accuracy. For $B=4$, accuracy is reduced and the correlation structure seen for $B=8$ or $16$ is not reproduced in Figure~\ref{fig:2D_histogram}a. 
This means that higher resolution of feature correlation, given by a bigger value of $B$, is more informative than higher dynamic range, which is obtained for smaller values e.g., $B=4$.

It could be assumed that the comparable accuracy between the raw and correlation trained models is explained by the capacity of the CNN to learn the feature correlation. However, the result shown in Section~\ref{sec:mgr} for late fusion suggests that the two representations are not exactly the same. 

Results do not confirm our assumption on feature robustness. However, our proposed method contributes in terms of feature dimensionality. Classifying impairment types in audio quality assessment application can benefit from having a reduced feature dimensionality for saving bandwidth in real time monitoring and storage in case of large volume of data. Given that the motivation of using the proposed method is related to feature robustness instead of feature dimensionality reduction, a comparison with other methods for the latter task is not given in this paper. Nevertheless, research studies on feature dimensionality reduction methods such as PCA and factor analysis, suggest that these methods require a large ratio between number of samples and number of features in order to preserve consistency in the data \cite{osborne2004sample, mundfrom2005minimum, shaukat2016impact}. This is crucial as the dataset that we used has an approximate ratio observations-features that is slightly lower than $2$:$1$ which is not the recommended ratio for using PCA or factor analysis. Therefore, as the proposed method is not dependent from the ratio observation-features, it could be preferred to the above-mentioned methods on datasets that show this peculiarity.

\section{Conclusions and Future Work}
In this paper we presented a new way to represent hand-crafted features for audio impairment recognition which is based on the correlation between features. We obtained significant performance improvement in terms of feature dimensionality and running time while maintaining accuracy levels similar to those obtained using raw feature representation. We also obtained a reduced feature dimensionality for music genre recognition i.e., $\approx30\%$ smaller, at the expense of a slightly lower accuracy.

In the future we want to explore different features to see if the method is robust in more scenarios. In particular, we want to go beyond the usage of low-level features by exploring the correlation-based representation on high-level features. We also intend to explore methods that further decrease the feature dimensionality and the runtime by compressing the sparse matrices obtained from the proposed representation. Finally, we believe that a comparison with other feature dimensionality reduction methods is needed. In particular, we want to compare the proposed method with other feature dimensionality reduction methods such as factor analysis and PCA using a dataset with higher ratio observations-features and with other feature selection methods such as correlation-based elimination and backward feature elimination.  

\bibliographystyle{IEEEtran.bst}
\bibliography{conference_101719.bib}

\end{document}